# Observation of a Zeeman Induced Lifshitz Transition in $URu_2Si_2$

B.S. Shivaram[a] and V.W. Ulrich[1],

Department of Physics

University of Virginia, Charlottesville, VA. 22901

and

D.G. Hinks

Materials Science Division, Argonne National Labs, Argonne, IL.60637.

## ABSTRACT

High resolution longitudinal sound velocity measurements in a magnetic field performed at $T \rightarrow 0$ in $URu_2Si_2$ reveal a second signature at a field $B_{LT}$ in addition to the step change in velocity expected at the superconducting upper critical field, $B_{c2}$. Characteristic Fermi surface related magneto-acoustic quantum oscillations (MAQO) emerge beyond the field $B_{LT}$ at a frequency ~160 T for B||a-axis. Measurements performed with B oriented at various angles between the a and c-axes reveal an anisotropy for $B_{LT}$ nearly identical to that of $B_{C2}$ suggesting similarity of the electron states involved in both the transitions. Given the observed frequency of ~160 T in the MAQO the transition is most likely related to the emergence of the $\eta$ Fermi surface in $URu_2Si_2$ (Lifshitz transition).

(a) To whom all correspondence should be sent - bss2d_AT_virginia.edu

[1] Current Address: Grove City College, Pennsylvania, Pa.-16127.

The heavy electron semimetal $URu_2Si_2$ has been well studied for almost three decades and yet continues to present us with many puzzling features. An outstanding issue has been the nature of the hidden order below T=17 K to explain which considerable effort has been devoted[1]. While the topic of hidden order is discussed mostly in the context of $URu_2Si_2$ a larger unsolved and related problem is the renormalization of the Fermi surface (FS) when the localized f-spins hybridize with the conduction electrons in a Kondo lattice. We have shed partial light on this issue in recent work where we show that a single site, single energy scale model is able to capture most of the experimentally observed features of a metamagnetic transition[2]. Such a transition seen in many of the Kondo lattice compounds is believed to be largely due to the reconstruction of the FS in high magnetic fields[3]. The success of the single site model contrasts with this interpretation. In $URu_2Si_2$ the FS has been probed repeatedly and is fairly well established[4,5,6,7,8,9,10,11]. Several changes in the FS occur leading up to a field of 35 T where metamagnetic behavior is observed[12]. Recent very low temperature Shubnikov-de Haas measurements have demonstrated[13] a pocket-by-pocket full polarization of the FS which occurs sequentially at 17 T, 24 T, ~30 T and finally at 35 T. It has been suggested that such small pocket polarization and field induced modifications of the FS are central to understanding the phenomenon of hidden order[13,14]. In the following we present detailed studies of the longitudinal sound velocity in $URu_2Si_2$ in the T $\rightarrow$ 0 limit. In high fields we show for the first time that there is a direct fingerprint of the FS evolution as seen through sound velocity or elastic constant changes. Although there have been previous ultrasound studies of $URu_2Si_2$ in high magnetic fields they did not cover the T $\rightarrow$ 0 limit[15,16,17].

The data presented here were obtained from measurements performed at the National High Magnetic Field Laboratory, Florida, both in a resistive magnet with fields upto 33 tesla and in a 20 T superconducting magnet with the sample immersed in the mix of the dilution refrigerator and mounted on a rotator. The sample used was the same as the one in which we have reported zero field longitudinal ultrasound velocity and attenuation measurements earlier[18] and was grown at Argonne National Labs in a vertical float zone refining furnace. A single transition at B=0 was observed in these studies. Our sample is of similar quality to that used in transport measurements by Kwok and coworkers[19]. In their work a RRR of 34 was measured.

In fig. 1 we show the measured change in the longitudinal sound velocity, Δv/v, at an operating frequency of 59.9 MHz, obtained as a function of the magnetic field applied parallel to the wavevector q which was also oriented along the a-axis. As seen in fig.1 there are clearly two step changes in the longitudinal sound velocity apparent only at the lowest temperatures.

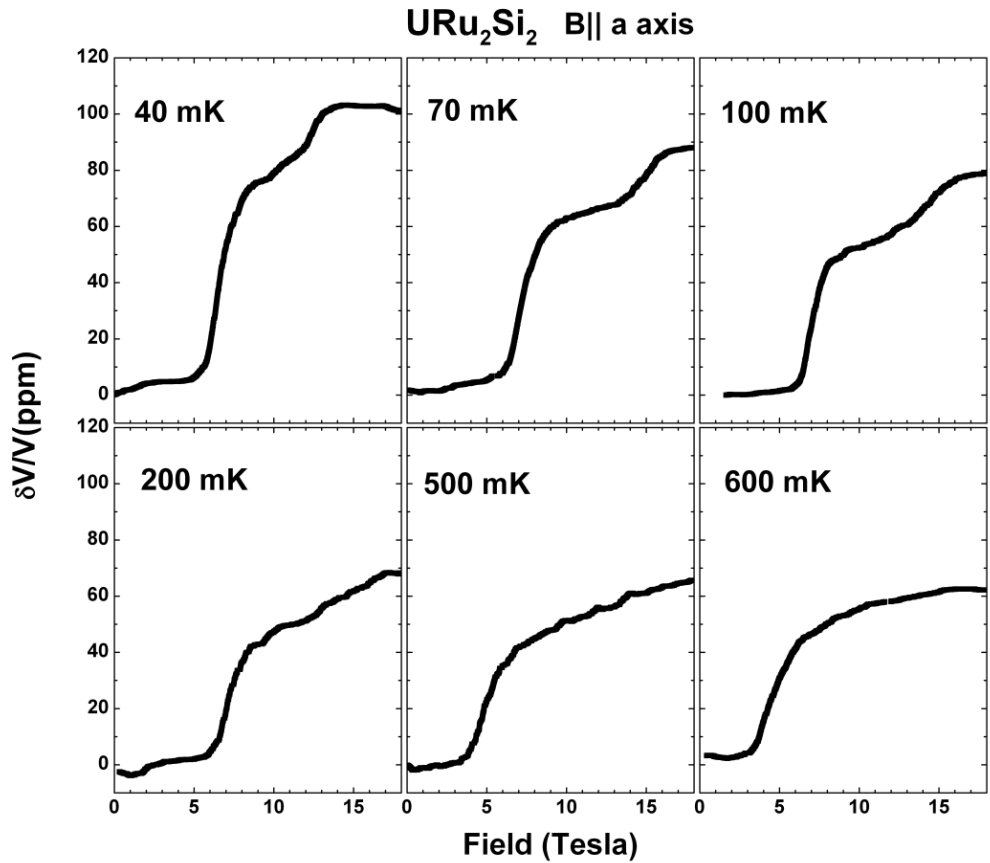

*Figure 1: The change in the longitudinal sound velocity in parts per million as a function of the magnetic field applied parallel to the a-axis in a single crystal of $URu_2Si_2$. The ultrasound was propagated with the wave vector parallel to the a-axis. The magnitude of the sound velocity jump at the lower transition is of the same order as the one observed in zero field at $T_c$ (ref.18).*

The lower field step corresponds to $B_{C2}$ and a second step at a higher field, $B_{LT}$, is also seen with its size weakening as the temperature is raised. Also apparent in this figure is the constancy of the sound velocity beyond $B_{LT}$. The observed magnitude of the jump in the sound velocity at $B_{c2}$ is comparable to the zero field value obtained at $T_c$ and reported by us previously[18]. This value is also in excellent agreement with estimates that can be made using thermodynamic identities. In a tetragonal crystal the change in the sound velocity along the a-axis is related to the elastic constant $C_{11}$, the jump in the heat capacity at $T_c$, $\Delta c_p(T_c)$, and the strain dependence of $T_c$ via the relation:

$$\frac{dv_{11}(T_c)}{v_{11}} = -\frac{\Delta c_p(T_c)}{2V_{mol}C_{11}(T_c)T_c}\left(\frac{dT_c}{d\varepsilon_1}\right)^2 \quad (1)$$

The strain dependence can in turn be estimated from reported uniaxial pressure dependence of $T_c$ via the relation:

$$\frac{dT_c}{d\varepsilon_1} = -(C_{11}\frac{dT_c}{dP_1} + C_{12}\frac{dT_c}{dP_2} + C_{13}\frac{dT_c}{dP_3}) \quad (2)$$

where the subscripts 1,2, and 3 refer to the three crystal axes a, b, and c respectively. For $URu_2Si_2$ $dT_c/dP_1= dT_c/dP_2 =-62$ mK/kbar, $dT_c/dP_3=43$ mK/kbar, and the heat capacity jump divided by $T_c$ is 54 mJ/mole $K^2$ and [20]. Using these values along with the value of $C_{11}=25.5 \times 10^{11}$ erg/cm$^3$ and $\rho=10.01$ gm/cm$^3$, (ref. 15) we obtain for the change in the longitudinal sound velocity for propagation along the a-axis, $\Delta v_{11}/v_{11}$ of 65 ppm the same as the value observed at the lower transition. Thus the jump in the sound velocity at the upper transition occurs in the normal metallic state of $URu_2Si_2$. It should be viewed as a stiffening of the lattice arising from a field induced transition restoring the FS. This stiffening occurs on top of that part which ensues as the FS is restored when Cooper pairs are destroyed at $B_{c2}$.

To precisely locate the positions of the two transitions, $B_{c2}$ and $B_{LT}$, we compute the derivative of the experimental velocity signatures. Representative derivative plots are shown in fig.2 where we define the positions of the two transitions as the field where the derivative peaks. The transitions admittedly are broad - this is not surprising since the sound velocity signatures reflect the properties of the bulk of the crystal and in $URu_2Si_2$ even the best samples show a distribution of $T_c$'s[21]. Nevertheless this does not prevent us from being able to locate the positions of the two signatures as both temperature and field orientation is varied.

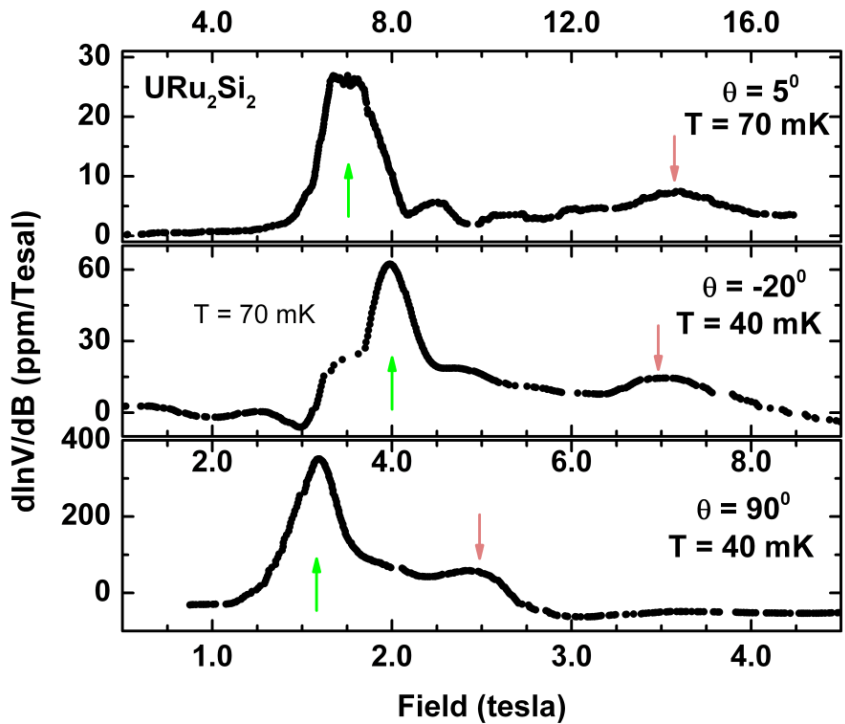

*Figure 2: Shows the derivative of the longitudinal ultrasound velocity change at three different orientations of the field with respect to the a-axis and at two separate temperatures. The position of $B_{c2}$ is identified by the green arrow (maximum in the derivative) and the upper transition $B_{LT}$ is marked by the red arrow. Note the change in scale for the horizontal axis for the top-most panel.*

Confining ourselves to the very low temperature results we plot in fig.3 the angular dependence of the positions of the two signatures. We reproduce the strong anisotropy of the upper critical field in the vicinity of B||a-axis well established by now in a number of studies[22]. Also notable in fig.3 is the angular anisotropy of the new transition, $B_{LT}$, which follows the same pattern and is almost a factor of two stronger than that of $B_{c2}$. Also shown in this figure are the values of $B_{C2}$ as obtained through a mutual inductance measurement performed on the same sample (seen supplementary information).

Further clues to this new transition are provided by examining our very low temperature data at the highest fields measured. Figure 4 is a plot of the sound velocity changes with an expanded vertical scale. It shows magnetoacoustic quantum oscillations that emerge beyond the new transition in the field range 20 T-33T. The measured frequency of these oscillations is

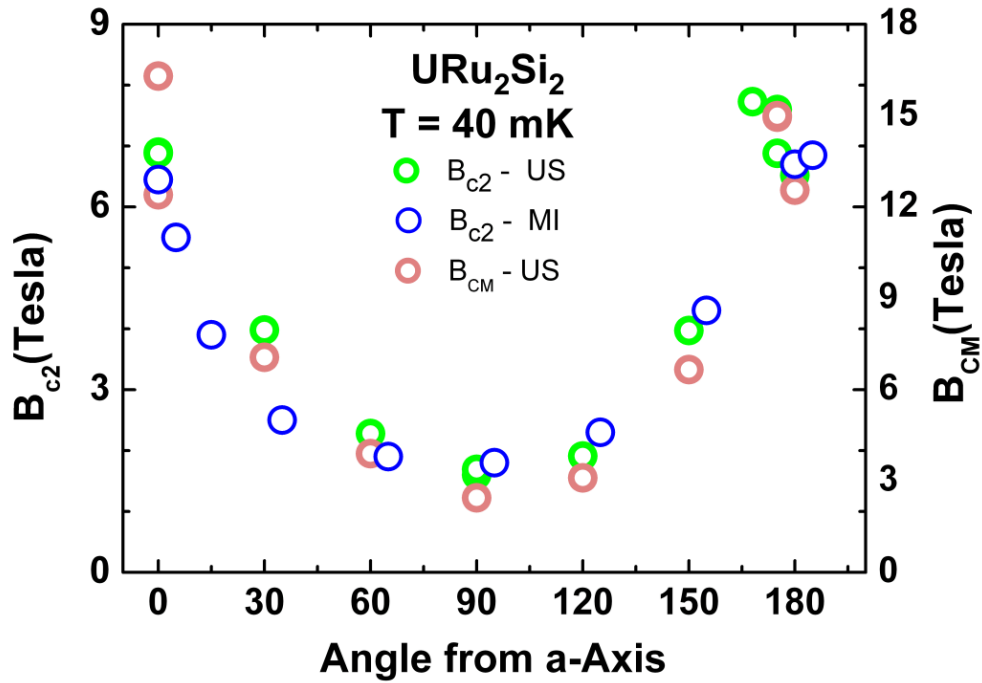

*Figure 3: Shows the measured values of the two critical fields (green circles - $B_{c2}$ and red circles - $B_{CM}$) as a function of angle with $90^0$ representing the c-axis. The blue open circles are the positions of the lower transition, $B_{c2}$, determined with a mutual inductance coil. The anisotropy in the upper transition is almost a factor of two larger than that of $B_{c2}$.*

approximately 160 T and is weakly field dependent. This frequency is in good agreement with an orbit for B||a-axis identified for the first time in the work of Okhuni et. al[23]. In subsequent work for B approximate to and along the c-axis frequencies closer to 90 T are observed and have been assigned to the η pocket of the FS[24]. In the absence of any other previously identified part of the Fermi surface in this frequency range we surmise that the lone point in fig. 13 of Okhuni et.al. and the oscillations observed in this work belong to the η pocket suggesting

that it is ellipsoidal with frequency increasing as the orientation of field is changed from c to a-axis .   With the available data set we are able to plot the position of this transition together with $B_{c2}$ for B||25 deg off a-axis and this is shown in the supplementary information.

Having associated the observed transition with a specific pocket of the FS we can turn to an interpretation of the observed stiffening of the lattice at $B_{LT}$ .  As argued earlier a natural explanation of the increase in the sound velocity at an electronic transition is that it be associated with the creation or "restoration" of the FS or part thereof.  Such a suggestion is not unreasonable since in $URu_2Si_2$ there is precedence for the sudden field induced appearance of a FS part.  In their Hall effect measurements Shishido et. al.[25] find an abrupt disappearance of the lower mass $\varepsilon$ orbit as the field is reduced even though other orbits associated with higher mass carriers survive .  This is a surprising result given that the area of the $\varepsilon$ orbit is larger than the neighboring $\alpha$ and $\beta$ parts of the FS.  Our results are similar - the MAQO attributable to the $\eta$ part of the FS appear suddenly as the field is increased.  The large increase in the sound velocity observed by us may be attributed to a singularity in the density of states that accompanies such a Lifshitz transition[26]. Such large changes in dv/v may also be understood within the context of the Lifshitz-Kosevich formalism employed to interpret quantum oscillations.  MAQO in the velocity are proportional to the fractional change in the FS cross section with strain and thus would be particularly strong for small cross sectional areas[27].  The transition at $B_{LT}$ being weakened at higher temperatures could arise from a thermal smearing in the density of states singularity.

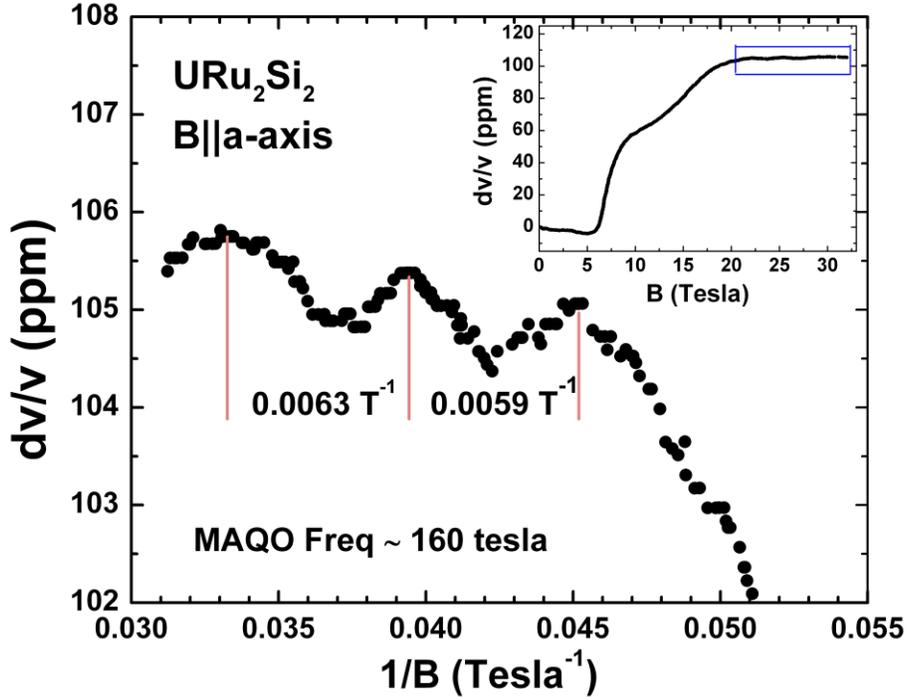

Fig.4: Shows the emergence of magneto-acoustic quantum oscillations in the longitudinal sound velocity above the new transition, $B_{LT}$ for T=100 mK. The observed frequency of these oscillations is consistent with the $\eta$ pocket of $URu_2Si_2$. The full dependence of the sound velocity starting from deep inside the superconducting state and including the region where the oscillations are observed (blue rectangle) is shown in the inset.

From the above discussion while the assignment of the observed large sound velocity changes to the $\eta$ pocket seems reasonable several open questions remain. As noted in ref.(13) there is a sequential polarization of many other parts of the FS as the field is brought to 35 T where the HO transition is eventually wiped out. Our measurements extended to 33 T and there is a notable absence of any additional step increase in the sound velocity in this field range. However, we note that other "suggestive signatures" in ultra low temperature measurements on $URu_2Si_2$ reported in the literature can be found. For example, in recent work Li and co-workers have carried out sensitive torque magnetometry measurements at very low temperatures (20 mK) and high magnetic fields (to 20T) and thus cover the same parameter range as in our work[28]. In particular their data for the field dependent magnetization for angle 16 deg away from the c-axis appears to indicate a distinct change in slope at 10 Tesla which is roughly a factor of two larger than $B_{c2}$ for this angle for their sample[29]. A change in slope of the magnetization is the second derivative of the free energy and a discontinuity implies a second order transition. From eqn.(1) assuming the strain dependence of the $B_{LT}$ transition is similar to that of the superconducting (SC) state using thermodynamic relations it may be surmised that a positive shift in the susceptibility $\chi$ will ensue.

In conclusion, we have presented field dependent high resolution sound velocity measurements in a zone refined single crystal of $URu_2Si_2$ which reveals a Zeeman induced Lifshitz transition. The transition exists only at the very lowest temperatures and disappears around 500 mK. The anisotropy of the field at which this transition occurs follows nearly the same angular dependence as $B_{C2}$. Our work calls for further very low temperature high field measurements such as magnetization and heat capacity which could also be influenced by the field induced singularity in the density of states accompanying a Lifshitz transition. Further magneto-acoustic measurements at lower temperatures to establish the effective mass of the electrons in the orbit observed here and angle dependent tracking of the MAQO are also called for.

Acknowledgements: We acknowledge the gracious help and encouragement from Drs. Eric Palm and Tim Murphy at the NHFML in Tallahassee where the measurements presented here were performed. The work at the University of Virginia was supported by the National Science Foundation through grant DMR 0073456. BSS acknowledges support from Argonne National Labs for a recent sabbatical appointment. The National High Field Magnetic Laboratory is supported by the National Science Foundation. We are most grateful to John Ketterson for many enlightening discussions. We acknowledge help from Vittorio Celli, Pradeep Kumar, Brian Maple and Michael Norman.

# Possible Identification of a Field Induced Spin Density Wave State in the 3D Semimetal $URu_2Si_2$

## Supplementary Information

The upper transition $B_{LT}$ is smeared out at higher temperatures. The following figure shows the location of the upper transition in a field-temperature phase diagram for an angle 25 deg. off a-axis.

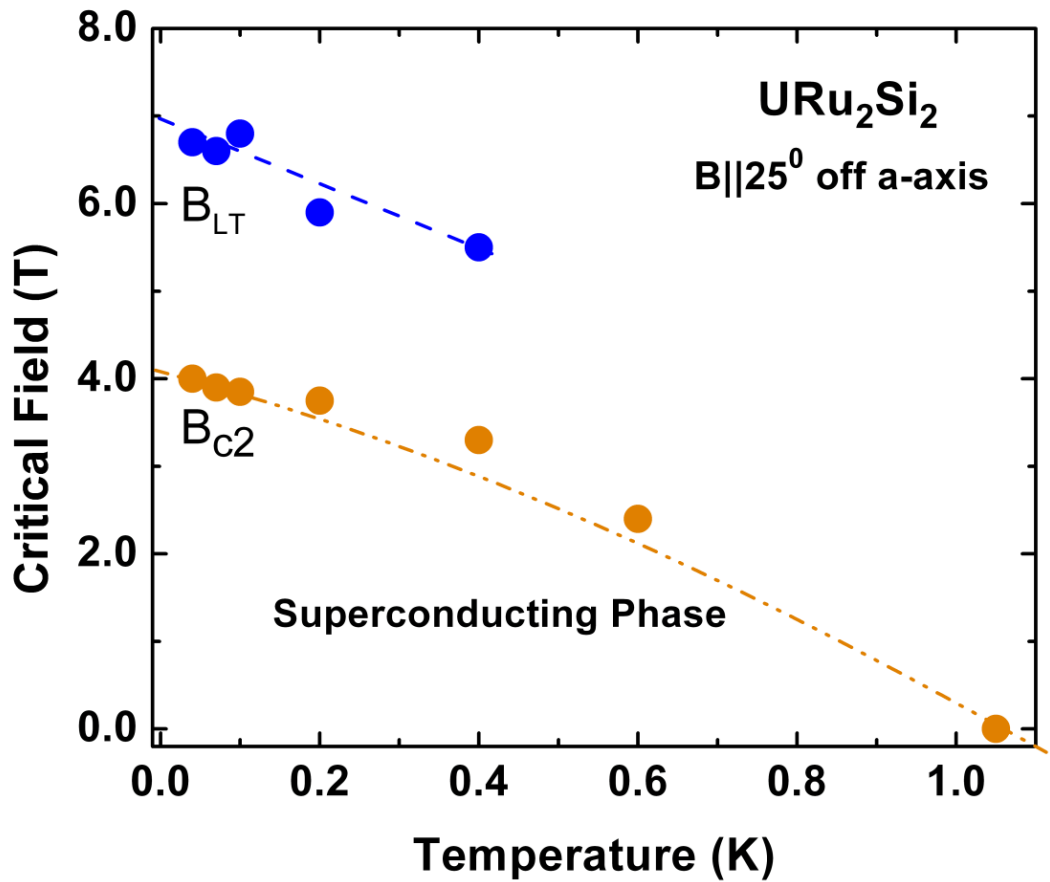

*Figure 1S: Shows the proposed off axis phase diagram for $URu_2Si_2$ constructed from the observed longitudinal sound velocity jumps.*

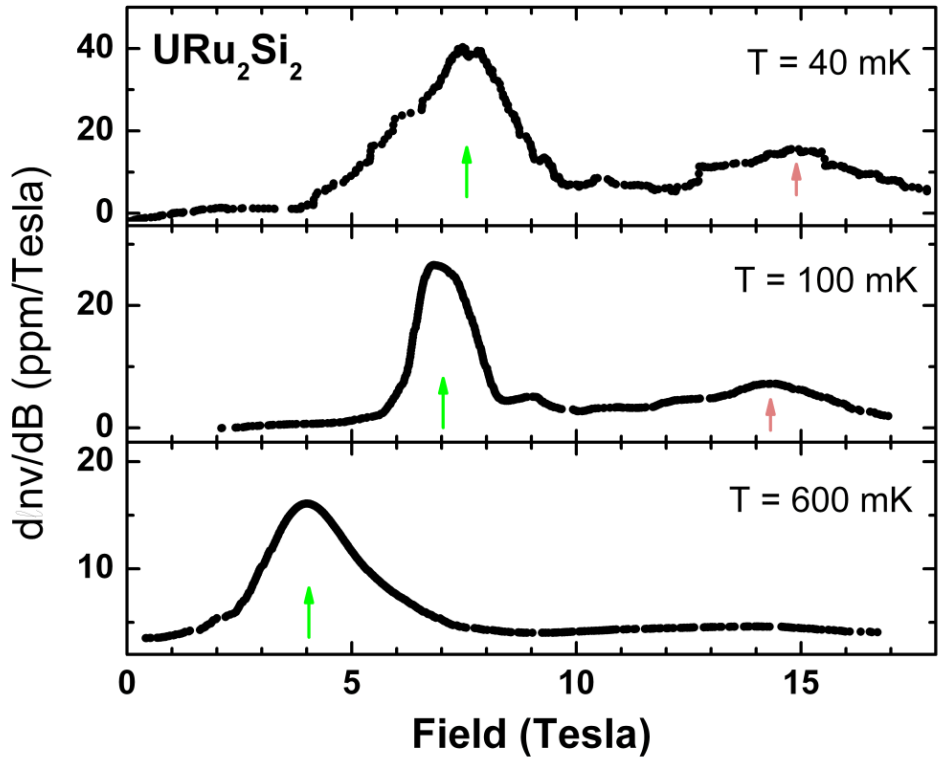

*Figure 2S: Additional representative plots of the derivative of the sound velocity showing the position of the upper critical field $B_{c2}$ (green arrow) and $B_{LT}$ the upper transition (red arrow). The orientation is 0 deg.*

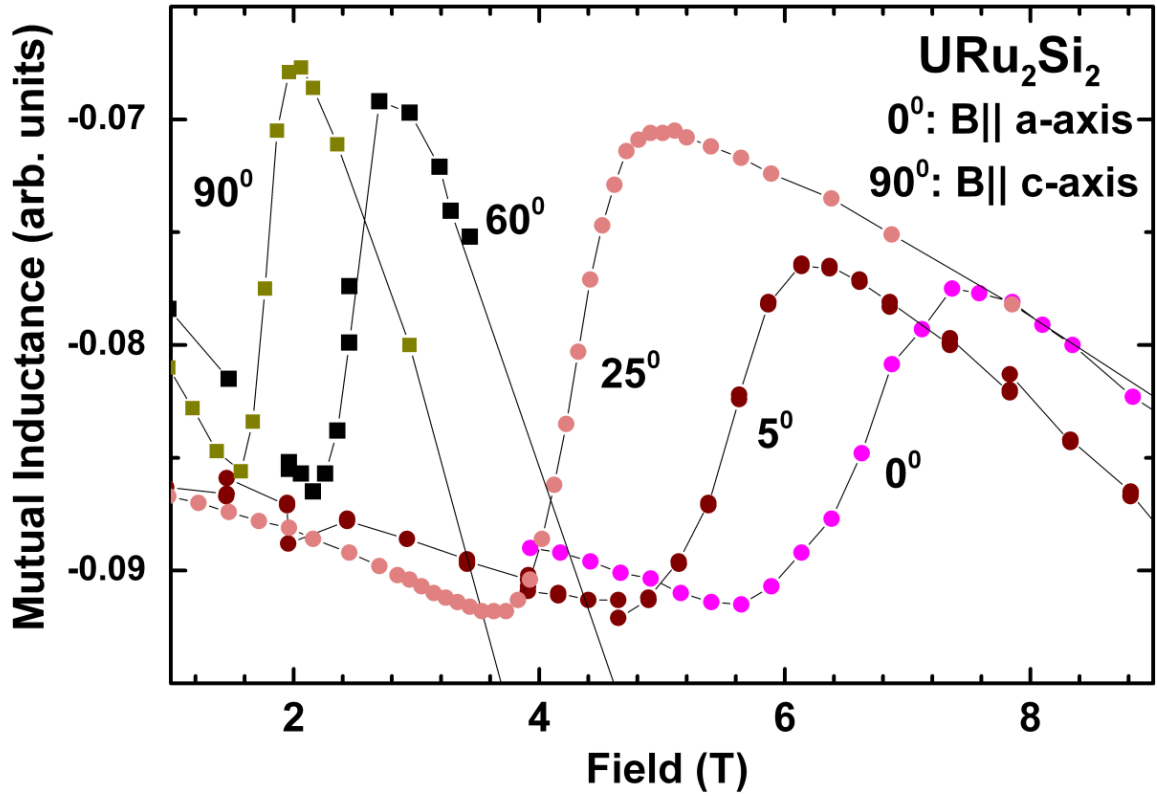

*Figure 3S: Shows the mutual inductance signals indicating the upper critical field $B_{c2}$ obtained for various angular orientations of the field with respect to the a-axis of $URu_2Si_2$*